\documentclass[12pt,preprint]{aastex}

\begin{document}

\title{The Integral Field Unit for the Echellette Spectrograph and Imager at Keck II}

\author{A. I. Sheinis \altaffilmark{1} \altaffilmark{2} }

\altaffiltext{1} {Astronomy Department, University
of Wisconsin, Madison, 475 N. Charter Street, Madison WI 53706} 
\altaffiltext{2} {formerly at Lick Observatory, Santa Cruz CA 95064}

\begin{abstract} We report on the design, development and commissioning of an Integral Field Unit (IFU) that has been built for the Echellette Spectrograph and Imager (ESI) at the W.M. Keck Observatory. This image slicer--based IFU, which was commissioned in the spring of 2004 covers a contiguous field of 5.65 x 4.0 arcseconds in 5 slices that are 1.13 arcseconds wide.  The IFU passes a spectral range of 0.39-1.1 um with a throughput of between 45 \% and 60 \% depending on wavelength and field position. The IFU head resides in an ESI slit mask holder, so that ESI may be converted to the IFU mode remotely by selecting the appropriate slit mask position. This IFU is the first of a family of designs for the spectrograph, providing a range of field-coverages and dispersions.   In addition, we present the first-light science imaging and spectroscopic observations of RXJ1131-123, a low-redshift, lensed quasar.  These observations show the 4 spectra of the lens and lensed-images captured in a single pointing.

\end{abstract}


\section{INTRODUCTION } \label{sect:intro}

An image slicer-based Integral Field Unit (IFU) has been developed for the Echellette Spectrograph and Imager (ESI) \cite{sbekmrbs02}. The IFU is an optical system that remaps a 2-dimensional field of view (FOV) onto the 20 arc-second spectrograph slit while preserving the F/\#, pupil location, and focal position of the telescope. ESI is a multi-mode Cassegrain-mounted spectrograph developed by the Lick Observatory Instrument Lab for the Keck II Observatory. It has been in regular operation since the first semester of 2000. ESI provides a reciprocal dispersion, up to R=13000 (2 pixels) with a wavelength coverage of 0.39 to 1.1 microns in a single exposure over a 20 arc-second slit.  Maximum throughput of the spectrograph in this mode has been measured at 29 \% , including atmospheric, detector and telescope losses. In this paper we discuss the commissioning, performance and first light science for the IFU-head addition to ESI. 

ESI was designed for very high throughput along with full visible spectrum coverage at moderate resolution.  In order to more fully utilize these capabilities a design study was undertaken from 1998 to 1999 to evaluate the possibility of modifying ESI for use as an integral field spectrograph.  That design study evaluated 3 IFU technologies for ESI:  Fiberoptic-based IFU such as the GMOS IFU at Gemini \cite{AMCDDMJHCM02} , lenslet-based systems such as Tiger and OASIS on CFHT \cite{BABCDDEFGMPRS95} and Sauron on WHT \cite{BCMMABMDEKPVZ01} and  slicer-based IFUÕs such as SPIFFI \cite{TTEMRB00}.  The conclusion of this study was that slicer-based technology was the most viable approach for the echellette mode of ESI.  This resulted in the funding of a prototype image slicer for ESI by the Keck Science Steering committee in 2000.  The prototype was finished in 2003 and tested on the mountain in early 2004.  This paper describes the development and testing and first light results of that prototype.

\begin{table}
\begin{center}
\caption{IFU Design Specifications}

\vskip 8pt
\begin{tabular}{lllll}
\hline\hline
\multicolumn{1}{c}{XDimension}  & YDimension &Reciprocal &number&slit width \\
(arcsec) & (arcsec) & Dispersion & of slices	&  (arcsec) \\
\hline
$4.0$	&	$5.7$	&	$3500$	&	$5$	&	$1.13$  \\
$4.0$	&	$3.75$	&	$5200$	&	$5$	&	$0.75$  \\
$4.0$	&	$2.5$	&	$7800$	&	$5$	&	$0.5$  \\
$4.0$	&	$2.1$	&	$13000$	&	$7$	&	$0.3$  \\

\hline\hline
\end{tabular}
\end{center}

\end{table}
\section{Specifications}

In order to create the most versatile instrument possible,  a series of IFU heads were designed.  Each head could be placed in one of 5 slit-mask bays of ESI, such that several different IFU observing modes could be selected remotely in the same way as the other observing modes of ESI.  Each IFU head has physical dimensions 150 X 80 X 60 mm, which is small enough that two IFUs may reside in one of the ESI slit mask bays . Presented in Table 1 are the four different IFU designs, each with a different FOV coverage and effective slit width and dispersion:  The design that was fabricated is the first entry in the table. The IFU optical designs are based on the advanced-image-slicer concept of \cite{C1997} .  

The fabricated IFU maps a contiguous 4.0 X 5.65 arc-second FOV onto the 20 arc-second echellette slit at a reciprocal dispersion of 3500. This reciprocal dispersion is sufficient for OH line rejection. The mechanical design for the system, described in \cite{SLHM02} is a substantial departure from existing systems. It is a completely monolithic, passive  and modular  implementation as an IFU head, (Figure 1).  Each IFU head resides in an ESI slit mask holder and is completely selectable as an observing mode.\\
\\
{\bfseries
IFU Specifications:
\begin{enumerate} 
\item{Wavelength range: 0.39-1.1 microns}
\item{Field of View: 4.0 x 5.65 arc seconds}
\item{Throughput:  45\% - 55\% depending on field and wavelength}
\item{Reciprocal Dispersion:  3500}
\item{Spatial Resolution: 0.3 x 1.13 arcseconds depending on seeing}
\item{Field of View: 4.0 x 5.65 Arc seconds}
\end{enumerate} 
}

\section{Optical Design}

The optical design is based on the Advanced-Image-Slicer concept of  \cite{C1997} the details of which are described in the figure 2 caption.  It has been optimized and analyzed in detail with the Zemax lens design program. This design uses all spherical imaging optics except for the final array of toroidal output slitlet mirrors.  These last optical elements in the IFU train act as the field lens to project the Keck pupil to the correct location within the spectrograph.  The optical design is such that the quality of the optical image from the telescope is not impacted by the IFU.  The optical design performance, shown in figure 1 demonstrates that the imaging spot sizes are well below that of the telescope.  

The optical layout is pictured in figure 2.  The optical system for this IFU is actually glued to the back-side of a spherical guide mirror.  A hole in the center of the mirror defines the approximate FOV of the IFU.

An in-house-developed dielectric-enhanced sliver coating has been applied to all optics. This coating has been optimized for the band-pass of the system. Measured throughput of the IFU ranges from 45  - 55 \% for the 7 reflections within the IFU.  Modeled performance predicted approximately 88 \% throughput.  It is believed that majority of the difference corresponds to a defect in the optical coatings on the IFU that resulted in a reflectivity of approximately 93-94 \% per surface as opposed to the 98 \%expected.  The remaining loss in throughput can be attributed to a slight vignetting within the IFU. Unfortunately it is not possible to recoat this device without complete disassembly, due to the geometry of the optical system.

\begin{figure}

\includegraphics[width=12cm,height=20cm,angle=0,keepaspectratio=true]{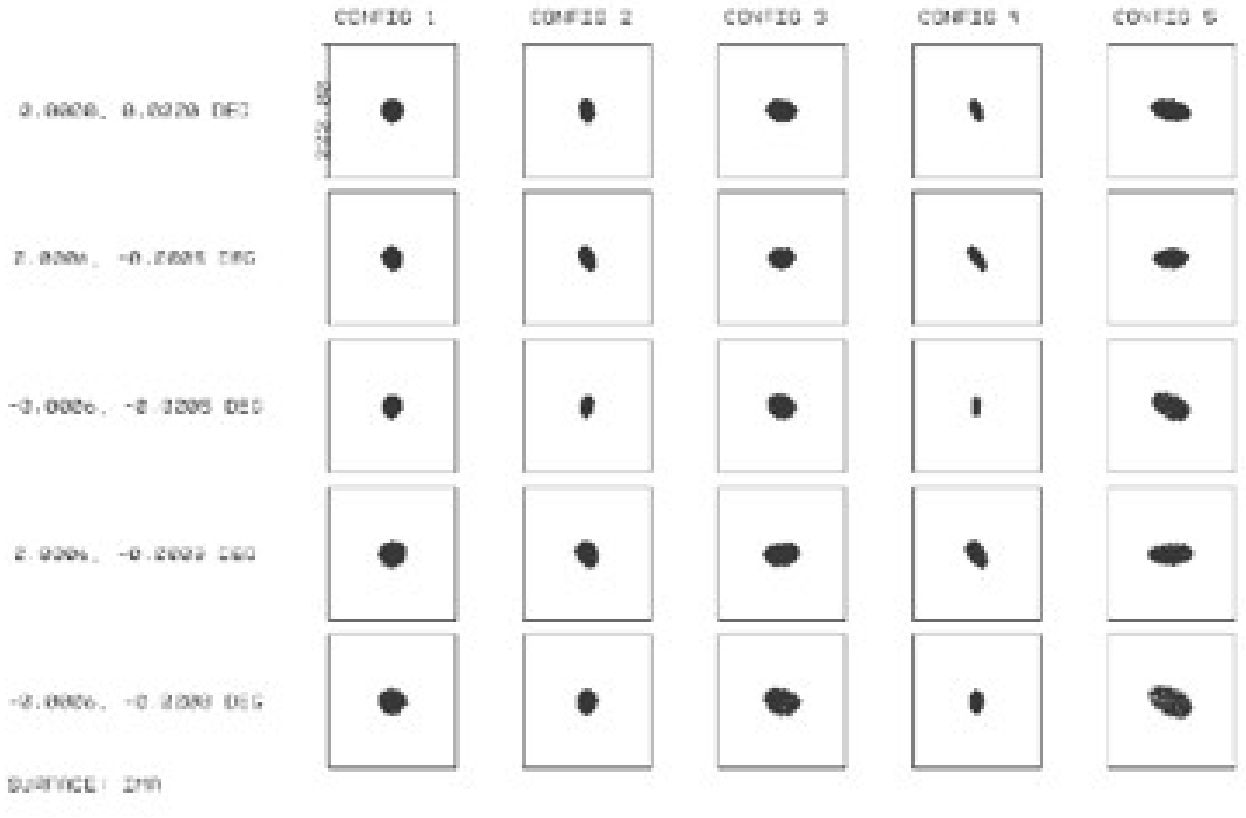}
\caption{Spot diagram at the full-field for all 5 slicer-paths. Each box is 1/10 arcsecond.. The optical performance of the slicer does not impact the image quality of the telescope or spectrograph.}
\end{figure}

\begin{figure}

\includegraphics[width=12cm,height=10cm,angle=0,keepaspectratio=true]{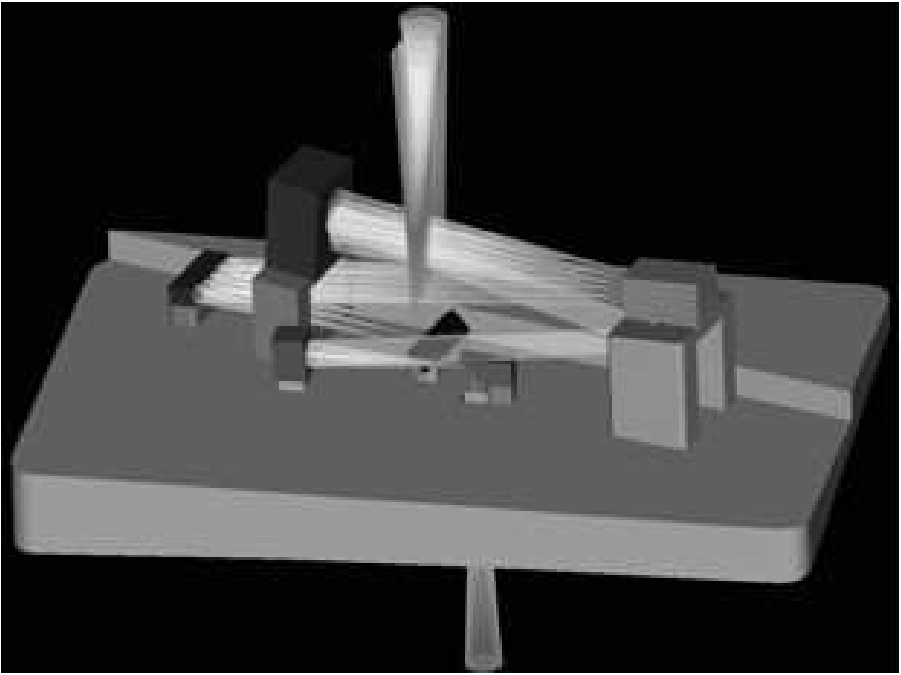}
\caption{The opto-mechanical layout of the system. From the bottom of the figure, the F/15 beam from the telescope  is shown coming  through the hole in the guider mirror and then to focus  at the Cassegrain image plane. A small pick-off mirror sends this beam to a three-mirror relay, which slows the beam to F/30 , remaps the 2-dimensional image onto  the image slicer (rightmost optic) element and images the Keck pupil onto an internal  pupil mirror array (leftmost optic) in the IFU train. From there, each slice of the image is sent to itÕs own two mirror relay, consisting of a spherical pupil mirror (leftmost optic) and a toroidal slitlet mirror. These two optics combine to remap the image onto a slit configuration, project the beam into the spectrograph (top of figure) and form a virtual image of the keck pupil at its original location.  Thus the beam entering the spectrograph appears to emanate from the Keck pupil, and has the image-plane shape of a single echellette slit.}
\end{figure}

\begin{figure}
\includegraphics[width=12cm,height=10cm,angle=0,keepaspectratio=true]{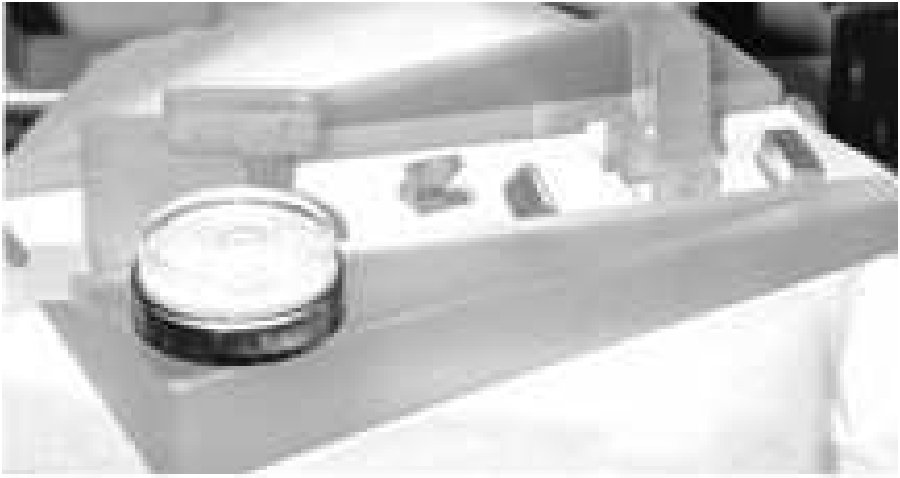}

\caption{The IFU final assembly of the IFU.  This view shows the IFU  mounted in the slit-mask interface carriage. Shown in this view, as seen from the telescope, is the guide mirror which serves as the substrate for the entire IFU assembly.}
\end{figure}

\begin{figure}
\includegraphics[width=12cm,height=10cm,angle=0,keepaspectratio=true]{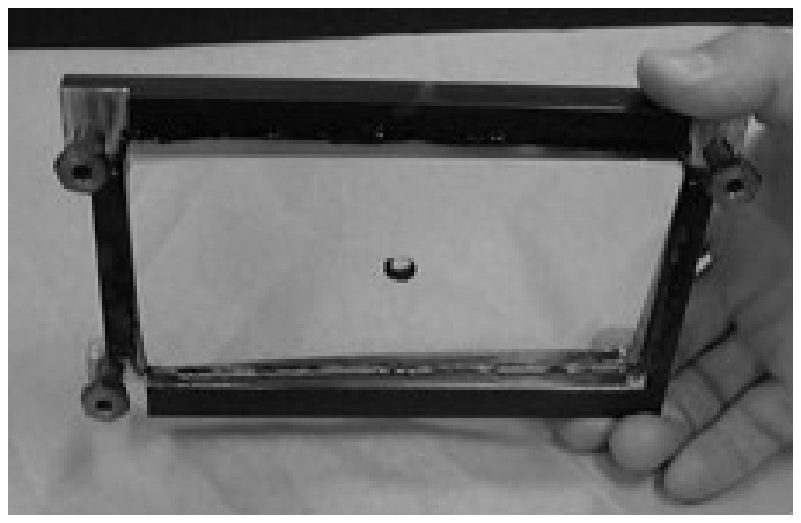}

\caption{The IFU final assembly.  This view shows the IFU  assembly prior to mounting in the slit-mask interface carriage. Shown in this view, as seen from the spectrograph, are the individual component optics mounted to the substrate.  Note, the bubble level is of order the size of a half dollar.}
\end{figure}

\section{Development}

The optical components for the IFU were fabricated and well the entire unit was assembled and tested at the Lick Observatory Instrument shops.  The finished product is shown in figure 3.  The device is all-Zerodur and has been demonstrated to maintain itÕs optical performance under extreme temperature variation, with some of the components tested at cryogenic temperatures.  


\section{ESI Modifications}

The only permanent modification to ESI was for added clearance forward of the telescope focus.   In order to access the Cassegrain focus the IFU was installed in the slit wheel.  The slit wheel was designed with approximately 25 mm of clearance in front of focus, while the IFU mechanics occupy approximately 40 mm in front of focus.  In order to accommodate this additional space requirement the front baffle plate within ESI had to be moved forward by approximately 20mm. This modification had no effect on any of the existing operational modes of ESI.

In addition, the departures from the standard observing modes are as follows:
\begin{enumerate} 

\item{The collimator requires a refocus as the output image plane resides approximately 7 mm below the Cassegrain focus.  This is easily accommodated through the instrument control interface with a change in collimator focal position, corresponding to collimator-encoder focus setting of  -19,500.}

 \item{This guider requires a refocus to raw position = -1900.} 
 
 \item{Slit wheel location requires an offset:  Since the input axis and output axis of the IFU vary by 20 mm, the slit wheel requires an offset from the nominal position.  This change is easily accommodated through the instrument control interface by offsetting the encoded position on the slit wheel, corresponding to a slit wheel encoder setting of 54000.}

\item{The telescope requires a refocus:  The IFU entrance image plane is located slightly in front of the ESI slit plane.  This corresponds to a telescope raw focus position of 0.00052212.}

 \end{enumerate}
 
 \section{Optical Performance}
 
 The optical throughput of the IFU head has been measured by recording the number of photons per second seen from spectrophotometric standard Feige 34 with the IFU head in place and measuring the number of photons seen from spectrophotometric standard Feige 98 without the IFU in place, i.e. measured though ESI in itÕs standard configuration with the 6 arcseconds wide-slit in place.

Each measurement has been corrected for atmospheric extinction and normalized to the expected number of photons for their respective spectrophotometric standard.  The resulting ratio of the throughput with the IFU to the throughput without the IFU for the middle 6 orders of ESI is plotted in figure 5.  The result is that the throughput attributed to the IFU alone is between 45 \% and 55 \% depending on wavelength and field position.
 
 \begin{figure}
\includegraphics[scale=0.8]{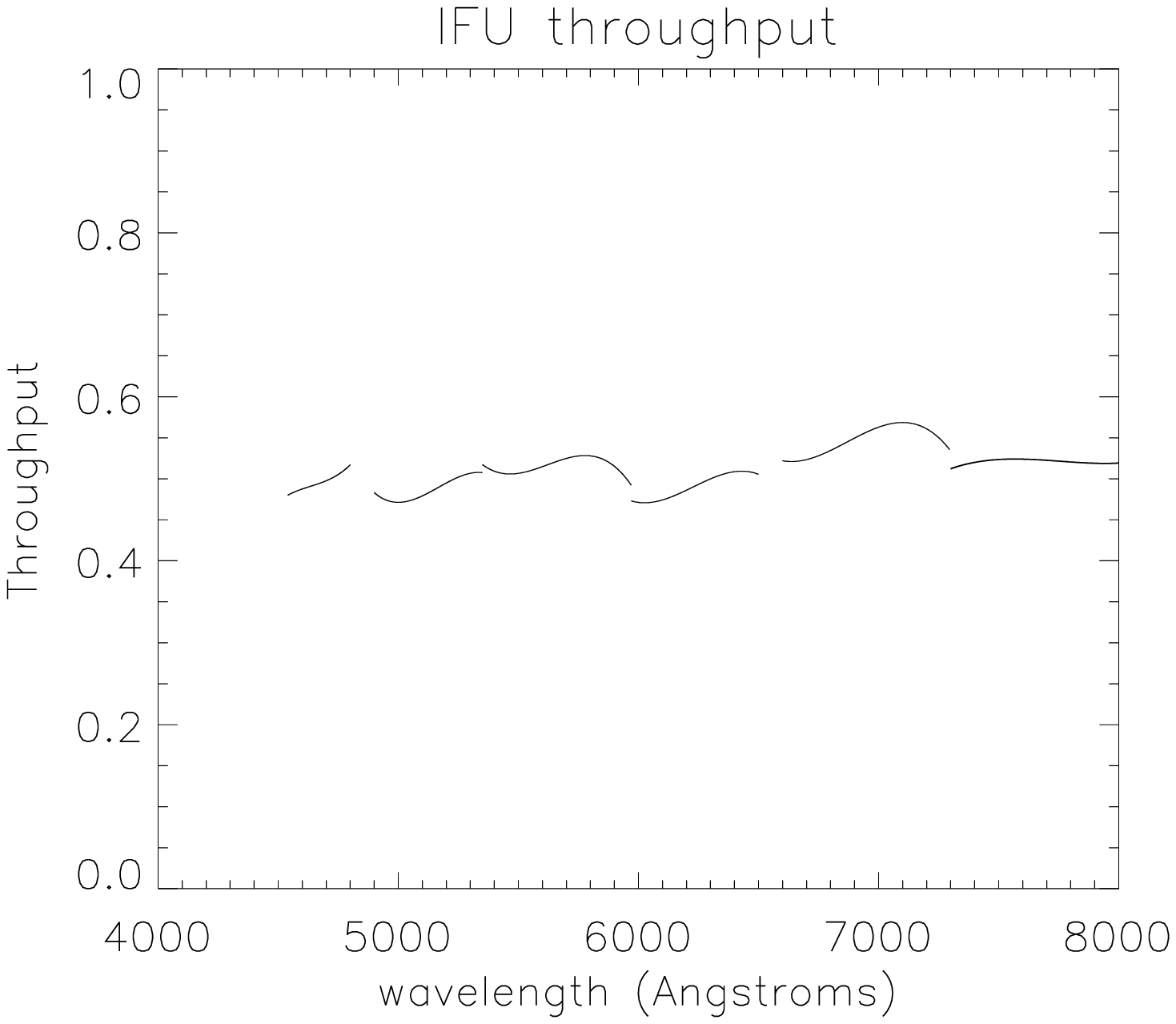}

\caption{Throughput of the IFU.  The ratio of photons incident on the IFU to photons detected divided by the ratio of photons incident on ESI without the IFU in place to those detected..}
\end{figure}

 \section{Scattered Light Rejection}ESI exhibits a small but non-negligible amount of scattered light in the echellette mode: of-order a few percent.  This scattered light shows up as a low-level undispersed intra-order flux, and must be modeled and removed from the between and under the echelle orders in order to get an accurate continuum flux level.  This is especially so in the case of observations of low surface brightness objects (LSBO) and even more so for LSBOÕs near bright sources i.e. off-nuclear observations of QSO host galaxies \cite{M&S02}.

The IFU has the added feature that it reduces the overall scattered light content by an order of magnitude.  This is because with the IFU in place there is no direct optical path into the spectrograph.  The IFU has 7 folds in the beam, each with an aperture and baffling, which serve to reject light that might enter the spectrograph from outside the science optical path.
 
 \section{Sky Subtraction}
 Unlike several other IFU devices built, the ESI IFU does not have a separate field aperture for sky-subtraction.  While a separate field would be useful, the 20 arcsecond slit length on ESI limits the total area available to IFU.  It was decided that breaking the entrance aperture up into a science and sky component would leave too small a field to be scientifically useful for many programs.

For small objects less than 4.0 x 5.65 arc seconds this does not present a problem as classical long-slit sky-subtraction techniques can be used.  In the case of large objects nodding techniques must be used.
 
 \section{First-Light Observations}
 In order to demonstrate the multiplexing capability provided by the IFU, we observed RXJ1131-1231 \cite{SSCHJCNBK03}, the lowest-redshift, lensed quasar currently known, during the commissioning run.  The goal was to simultaneously record the spectra of the 4 gravitationally lensed images of the quasar in a single pointing, in order to explore the substructure in the Dark Matter halo of the lensing galaxy by looking for variations in the magnification of the broad-line region (BLR) as compared to the narrow-line region (NLR) in several images of the same QSO \cite{MM03}.  
RXJ1131-1231 is part of the CASTLES survey of gravitationally lensed objects \cite{KFILMR99}.  It has a redshift of 0.658, with a redshift for the lens of 0.295.  V-band magnitude for the system is 16.7 with V=18.4 attributed to the lens.  The approximate diameter of the system is 3.69 arcseconds.   4 images were taken in R-band of 200 seconds each along with  6 spectroscopic images of 900 seconds each.  The co-added R-band images are shown in Figure 6, along with an overlay of the IFU field of view in the second panel. 
 
 The spectra of 3 of the 4 images along with the lensing galaxy were captured by the observation.  The 4th image was missed due to incomplete calibration of the IFU pointing angle at the time of first light. This issue has since been corrected.
The resulting co-added spectral images are shown in Figure 7.  This figure is unprocessed except for cosmic ray removal, bias removal and removal of the intrinsic anamorphic distortion in the instrument.  The 10 orders covering 0.39 to 1.1 microns are shown split into 5 different 2-dimensional spectra from the different spatial locations associated with the different slices.
 
 Figure 8 shows the extracted spectra for the 3 images captured with a single pointing.  The spectra have been flat-fielded, bias-subtracted and sky-subtracted in the normal way.  The extractions all are from order 8 and have been relative-fluxed to an arbitrary constant in order to show the relative flux ratios of the narrow forbidden lines of [OIII] to the broad permitted lines of $H_{\beta}$ for the three images.  To within the uncertainty in the measurements these ratios are seen to be constant between the 3 different gravitationally-lensed images.  This observed lack of variation in the magnification of the images suggests very little substructure in the dark matter halo of the lens.  This result, however is somewhat inconclusive as the spectro-photometric errors are still being evaluated.

  \begin{figure}
\includegraphics[scale=1.2]{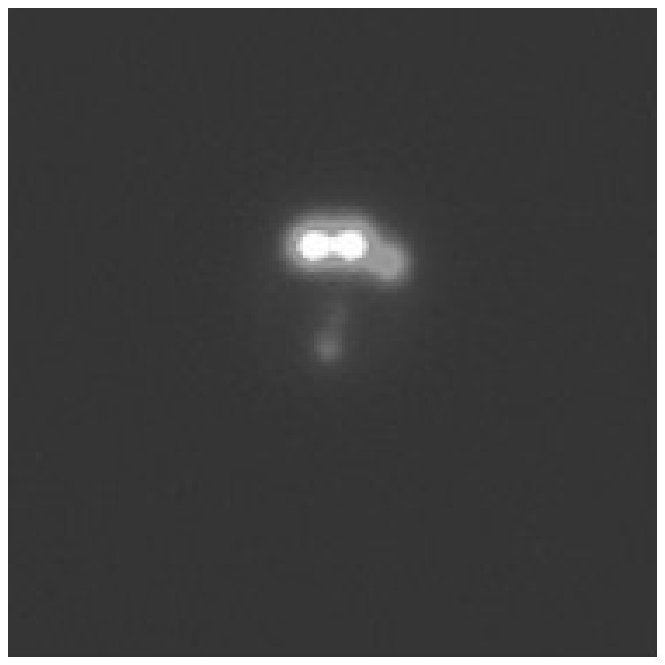}
\includegraphics[scale=1.2]{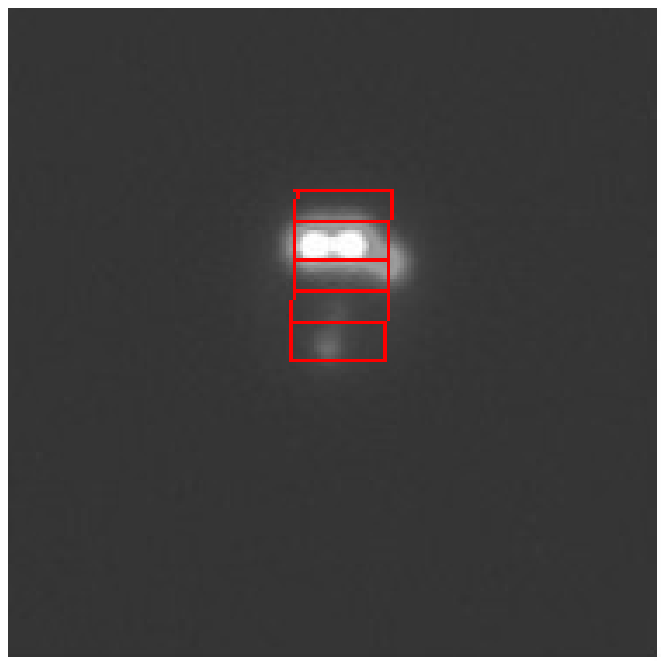}

\caption{Panel A shows the 4 x 200 second, R-band image of RXJ1131-1231, the lowest-redshift, lensed quasar currently known. Panel B shows the same image overlayed with a representation of the IFU field of view. The goal of the spectroscopic observation was to simultaneously record the spectra of the 4 gravitationally lensed images of the quasar in a single pointing.  3 of the 4 images were captured.  The 4th image was missed due to incomplete calibration of the IFU pointing angle at the time of first light. This issue has since been corrected.}
\end{figure}
 
 \begin{figure}
\includegraphics[scale=.74]{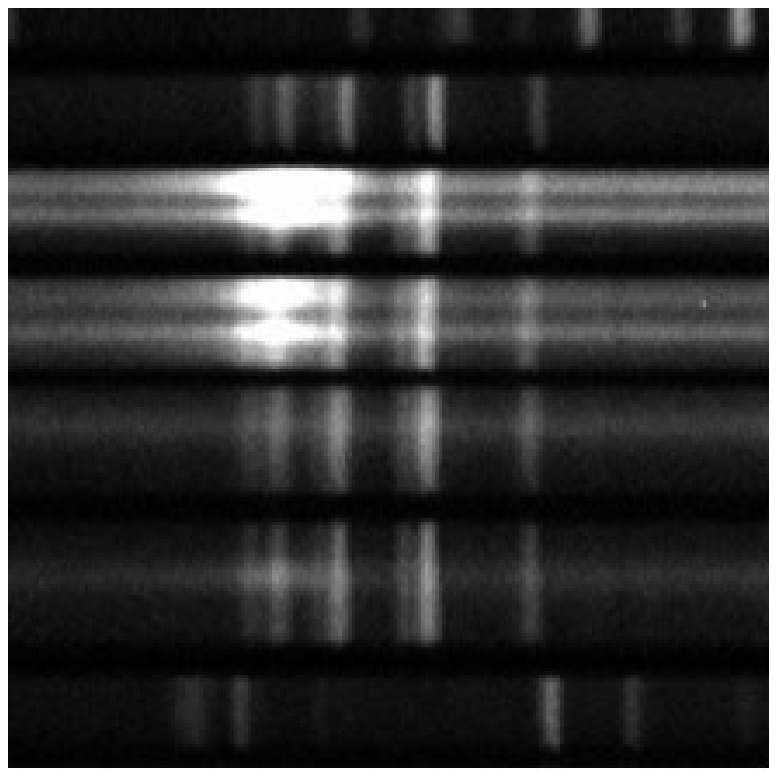}
\includegraphics[scale=1.0]{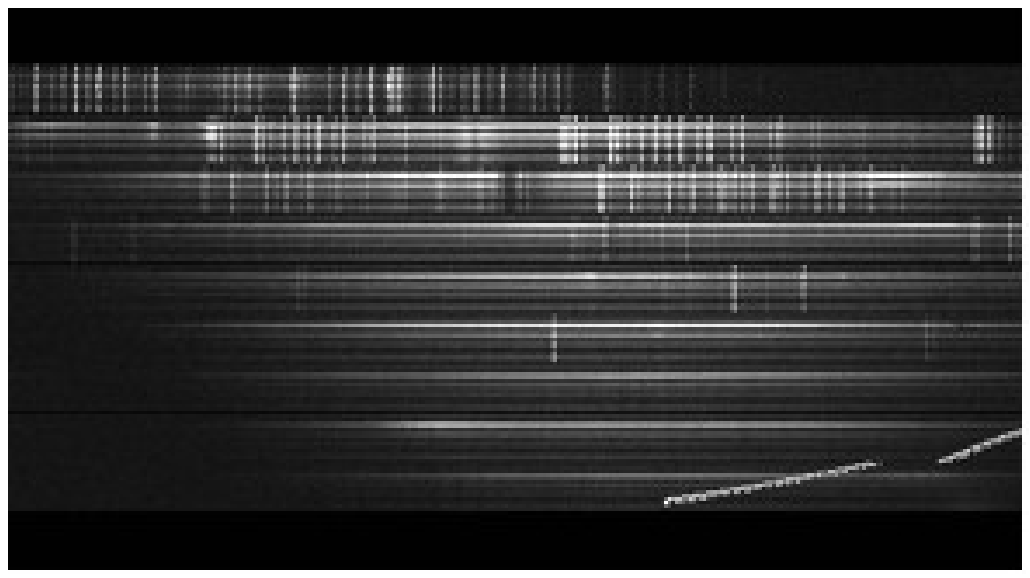}

\caption{shows the extracted spectra from RXJ1131-1231, the lowest-redshift, lensed quasar currently known taken during the commissioning run.  Panel A is the region about 5007 Angstroms in order 8, showing the 5 different 2 dimensional spectral from the different spatial locations associated with the different slices.  Panel B shows the full echellogram covering 10 orders from 0.39 to 1.0 microns.}
\end{figure}
 
 \begin{figure}
\includegraphics[scale=0.8]{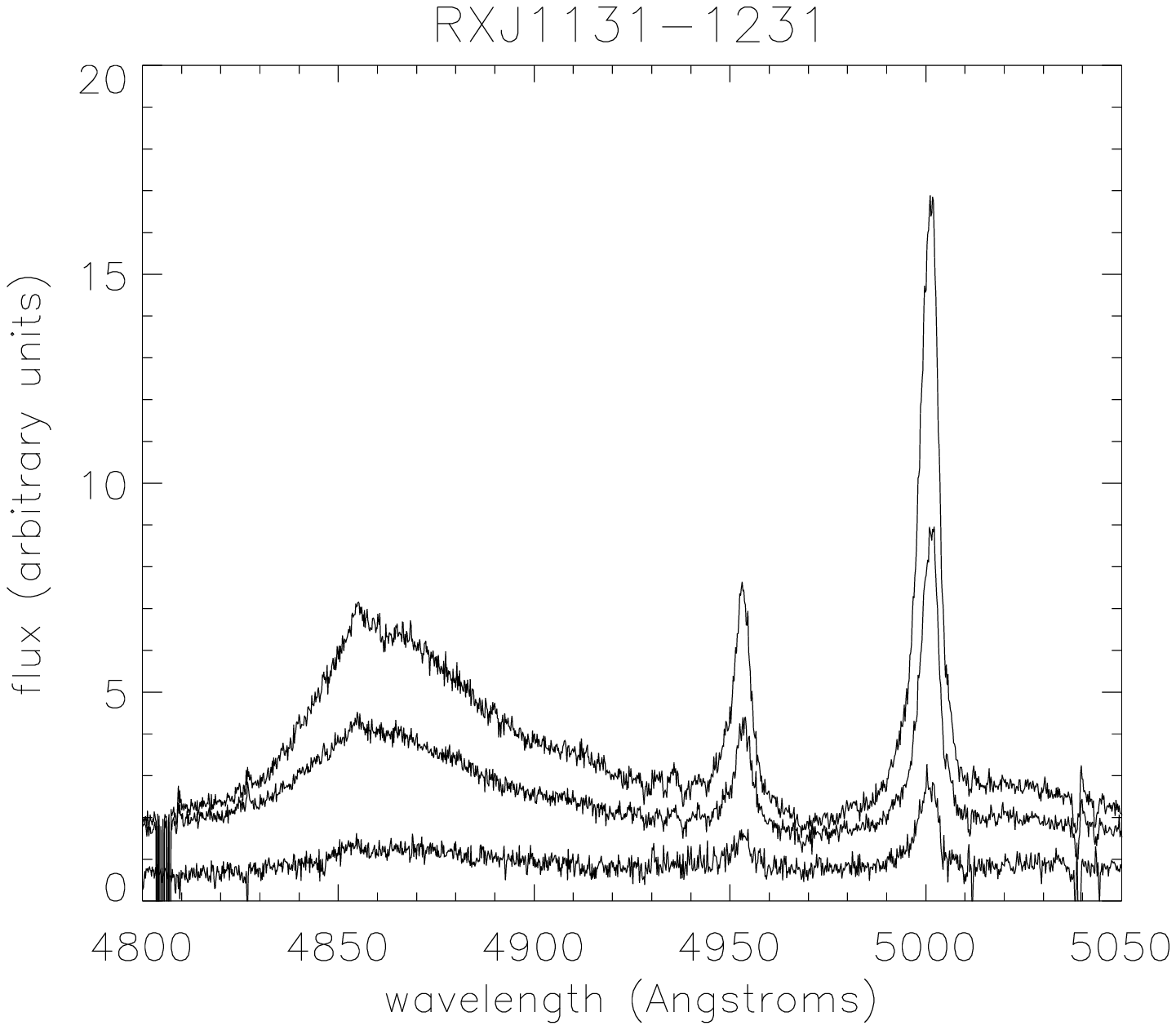}

\caption{shows the extracted spectra from RXJ1131-1231.  The three curves correspond to 3 of the 4 lensed images of the quasar, showing the relative fluxes of the H beta and the [OIII]4959 and 5007 lines.}
\end{figure}

 \section{Conclusions}
 
 We have demonstrated the performance of the first integral field unit to be commissioned at the Keck telescope.  This IFU is an add-on to the very-successful Echellette Spectrograph and Imager.  It allows ESI to cover a contiguous field of 5.65 x 4.0 arcseconds in 5 slices that are 1.13 arcseconds wide at a reciprocal dispersion of R=3500.  The IFU exhibited between 45 \% and 55 \% throughput while allowing the user to multiplex observations by up to a factor of 5.
 
 \section{Acknowledgments}
 
 The author would like to thank Joe Miller and Terry Mast for their very informative discussions and the staff at Keck Observatory, including Bob Goodrich and Jim Lyke for their help during the commissioning run. This project was funded by a grant from W.M. Keck Observatory and Sheinis was supported by an NSF Astronomy and Astrophysics Postdoctoral Fellowship under award AST-0201657.

 \bibliographystyle{apj}  

\end{document}